\documentclass[%
reprint,
superscriptaddress,
floatfix,
amsmath,amssymb,
prap,
]{revtex4-2}
\usepackage{graphicx}  
\usepackage{subfigure}
\usepackage{multirow}

\usepackage{siunitx}
\usepackage{fancyhdr}
\usepackage{longtable}
\usepackage{parskip}
\usepackage[T1]{fontenc}
\usepackage{dcolumn}   

\usepackage{bm}        
\usepackage{amsfonts}  
\usepackage{amsmath}   
\usepackage{amssymb}   
\usepackage{physics}

\newcommand{\pwisein}{\left\{ \begin{array}{ll}}
\newcommand{\pwiseout}{\end{array}\right.}

\usepackage{filecontents}

\begin{document}

\title{Parity and singlet-triplet high fidelity readout in a silicon double quantum dot at $0.5~\text{K}$}

\author{David J. Niegemann}
\email{david.niegemann@neel.cnrs.fr}
\affiliation{Univ. Grenoble Alpes, CNRS, Grenoble INP, Institut N\'eel, 38402 Grenoble, France}

\author{Victor El-Homsy}
\affiliation{Univ. Grenoble Alpes, CNRS, Grenoble INP, Institut N\'eel, 38402 Grenoble, France}

\author{Baptiste Jadot}
\affiliation{Univ. Grenoble Alpes, CNRS, Grenoble INP, Institut N\'eel, 38402 Grenoble, France}

\author{Martin Nurizzo}
\affiliation{Univ. Grenoble Alpes, CNRS, Grenoble INP, Institut N\'eel, 38402 Grenoble, France}

\author{Bruna Cardoso-Paz}
\affiliation{Univ. Grenoble Alpes, CNRS, Grenoble INP, Institut N\'eel, 38402 Grenoble, France}

\author{Emmanuel Chanrion}
\affiliation{Univ. Grenoble Alpes, CNRS, Grenoble INP, Institut N\'eel, 38402 Grenoble, France}

\author{Matthieu Dartiailh}
\affiliation{Univ. Grenoble Alpes, CNRS, Grenoble INP, Institut N\'eel, 38402 Grenoble, France}

\author{Bernhard Klemt}
\affiliation{Univ. Grenoble Alpes, CNRS, Grenoble INP, Institut N\'eel, 38402 Grenoble, France}

\author{Vivien Thiney}
\affiliation{Univ. Grenoble Alpes, CNRS, Grenoble INP, Institut N\'eel, 38402 Grenoble, France}

\author{Christopher B{\"a}uerle}
\affiliation{Univ. Grenoble Alpes, CNRS, Grenoble INP, Institut N\'eel, 38402 Grenoble, France}

\author{Pierre-Andr\'e Mortemousque}
\affiliation{CEA, LETI, Minatec Campus, F-38054 Grenoble, France}

\author{Benoit Bertrand}
\affiliation{CEA, LETI, Minatec Campus, F-38054 Grenoble, France}

\author{Heimanu Niebojewski}
\affiliation{CEA, LETI, Minatec Campus, F-38054 Grenoble, France}

\author{Maud Vinet}
\affiliation{CEA, LETI, Minatec Campus, F-38054 Grenoble, France}

\author{Franck Balestro}
\affiliation{Univ. Grenoble Alpes, CNRS, Grenoble INP, Institut N\'eel, 38402 Grenoble, France}

\author{Tristan Meunier}
\affiliation{Univ. Grenoble Alpes, CNRS, Grenoble INP, Institut N\'eel, 38402 Grenoble, France}

\author{Matias Urdampilleta	}
\email{matias.urdampilleta@neel.cnrs.fr}
\affiliation{Univ. Grenoble Alpes, CNRS, Grenoble INP, Institut N\'eel, 38402 Grenoble, France}



\date{\today}

\begin{abstract}  

Pauli spin blockade measurements achieved so far the highest fidelity of spin readout in semiconductor quantum dots,overcoming the 99\% threshold. Moreover, in contrast to energy selective readout, PSB is less error prone to thermal energy, an important feature for large scale architectures which will likely be operated at temperatures above a few $\SI{100}{\milli\kelvin}$. 
In this work, we use RF-reflectometry charge detection to readout the spin state of a double quantum dot. We demonstrate that it is possible to not only perform a standard singlet-triplet readout but also a parity measurement which allows to distinguish $T_0$ and the polarized triplets $T_-$ and $T_+$ states.
Moreover, we achieve high fidelity spin readout with an average fidelity above $\SI{99.9}{\percent}$ for a readout time of $\SI{20}{\micro\second}$ and $\SI{99}{\percent}$ for $\SI{4}{\micro\second}$ at a temperature of $\SI{0.5}{\kelvin}$. 
Finally, we succeed to initialize a singlet state in a single dot with a fidelity higher than $\SI{99}{\percent}$ and separate the two electrons while keeping the same spin state with a $\approx\SI{95.6}{\percent}$ fidelity.

\end{abstract}

\maketitle 

\section{Introduction}

The system size of today's semiconductor quantum dots remains in the few qubits regime \cite{Philips_2022, Hendrickx_2021}, but the community already works on scalable designs for qubit processors\cite{Veldhorst_2017}. Scaling up qubit systems goes along with an increased interest in cointegration of control electronics\cite{Ruffino_2021,Xue_2021}. This would result in higher power dissipation at the quantum chip level and the necessity to work at elevated temperatures (beyond dilution fridge base temperature), where the cooling power is typically in the hundreds of $\SI{}{\milli\watt}$ range \cite{Urdampilleta_2019}. Important progresses in this direction have been made in particular the realisation of high fidelity single qubit gate and two-qubit gate \cite{Yang_2020, Camenzind_2022, Petit_2020} above $\SI{1}{\kelvin}$. However, spin readout fidelity and initialisation is often a limiting process due to thermal broadening of reservoirs or the presence of low lying excited valley state.
In this context, the three steps of qubit operation, namely initialization, manipulation, and readout, all need to be fast compared to the decoherence rate in order to perform error correction protocols. For Si spin qubits, this requires $\SI{}{\mega\hertz}$ readout frequency with a fidelity above the $\SI{99.9}{\percent}$ threshold to ensure that readout is not the bottleneck in the operation of a quantum processor. Additionally, the readout should ideally come with a small footprint and gate overhead, enabling large scale architectures.\\
Typical spin readout in quantum dots requires a spin-to-charge conversion mechanism. Common techniques are energy selective readout \cite{Elzerman_2004} and Pauli spin blockade (PSB) \cite{Urdampilleta_2019,Lai_2011, Bohuslavskyi_2016}. Pauli spin blockade showed so far the highest readout fidelity, achieving fidelities $>\SI{99}{\percent}$\cite{Borjans_2021, Blumoff_2022}. Moreover, PSB does not require a nearby reservoir and was demonstrated at temperatures as high as $\SI{4.5}{\kelvin}$ \cite{Camenzind_2022}. PSB requires two spins, giving rise to one singlet and three triplet states. As PSB just provides one bit of information, further readout is required to determine the system state completely. Two different Pauli spin blockade readouts have been observed \cite{Seedhouse_2021}. The so-called ST-readout allows to distinguish the singlet $S_0$ state from the three triplet $T_-, T_0,$ and $T_+$. Fast $T_0$ relaxation, e.g. through $S_0/T_0$ mixing, can yield to the same signal as $S_0$. This so called parity readout allows to distinguish the polarized spin states from the non-polarized ones. Using a combination of both of these readouts allows to distinguish $T_0$,$S_0$ and $T_- (T_+)$, a requirement for the full tomography of a two spin-1/2 system \cite{Rohling_2013}.\\ 
The charge sensor of choice in many devices is a single electron transistor or quantum point contact, with at least three electrodes\cite{House_2015}. This large gate overhead poses scalability challenges, especially if one wants to have local charge sensors in 2D-arrays \cite{Mortemousque_2021}. A solution is RF-reflectometry, not relying on a current measurement, but a capacitive measurement. This allows to reduce the readout device to a single gate electrode to form an ancillar quantum dot\cite{Chanrion_2020, Ansaloni_2020, Borjans_2021}.\\
In this work we combine scalable fabrication technology, PSB and RF-reflectometry to demonstrate rapid and high fidelity single shot readout of spins in a double quantum dot. We work at a temperature of $\SI{0.5}{\kelvin}$, use a single lead quantum dot as a charge sensor and perform PSB with an average fidelity and visibility exceeding $\SI{99.5}{\percent}$ in a device fabricated in a $\SI{300}{\milli\meter}$ foundry.

\section{Device fabrication and experimental setup}
The device used in this work, similar to the one depicted in \ref{fig:Panel1}(a), is fabricated on a 300-mm silicon-on-insulator substrate. A $\SI{80}{\nano\meter}$ wide silicon channel is defined by mesa patterning and is separated from the substrate by a buried oxide of $\SI{145}{\nano\meter}$. A $\SI{6}{\nano\meter}$ thermally grown SiO$_2$ is used to separate the gates from the nanowire. The gates are made by atomic layer deposition of TiN of $\SI{5}{\nano\meter}$ and $\SI{50}{\nano\meter}$ of poly-Si. A bilayer hard mask of $\SI{30}{\nano\meter}$ SiN and $\SI{25}{\nano\meter}$ SiO$_2$ is on top of the metallic gates. Using a hybrid deep-UV-electron-beam gate-patterning scheme allows to transfer the gate structure into the hard mask by an alternation of lithography and etch steps. A final etch step separates the gates, resulting in a total of 6 gates ($2\times 3$) of $\SI{40}{\nano\meter}$ in gate width, longitudinal and lateral and transverse spacing. Before doping of the reservoirs, Si$_3$N$_4$ spacers of $\SI{35}{\nano\meter}$ are deposited to cover the space between the gates. Then, the source and drain reservoirs, labeled $S$ and $D$ respectively, are n-type doped using ion implantation, followed by an activation step using an $N_2$ spike anneal. Finally, the device is encapsulated and the gates are connected to Al bond pads.\\
The device chip is glued to a PCB and all gates except the gate B2 are connected to DC wires. The gate B2 is connected to a DC wire and a coaxial-cable through a bias-T with a $\SI{20}{\kilo\hertz}$ DC cut-off. Connecting the source reservoir to a Nb spiral inductor with $L=\SI{69}{\nano\henry}$ forms together with the parasitic capacitance $C_p$ of the circuit an LC-resonance at $f_{res} = \SI{1.2}{\giga\hertz}$. We extract from the resonance frequency a parasitic capacitance of $C_p \approx \SI{0.25}{\pico\farad}$. The amplitude and phase of the reflection of an RF-tone close to the resonance of the LC-circuit changes if the resonance frequency of the circuit is changed. A shift of the resonance frequency occurs when the electrochemical potential $\mu$ of the reservoir is aligned with an energy level of a nearby quantum dot and electron tunneling can occur\cite{Cottet_2011, Gonzalez-Zalba_2015}. We use IQ demodulation to sense the signal change. Thus, source reflectometry allows sensing of the nearby quantum dots defined by gates T1 and B1 (see suppl. mat. A).\\
The device is cooled down in a dilution fridge with a variable temperature control. The present experiment is achieved at a base temperature of $\SI{500}{\milli\kelvin}$. At this temperature, applying a positive voltage on the gates results in the accumulation of quantum dots at the Si/SiO$_2$ interface. We can hence form an array of up to $2 \times 3$ quantum dots. In this paper we will only present data using the leftmost $2 \times 2$ array.

\section{Charge sensing using RF-reflectometry}
\begin{figure}[t]
\includegraphics[width=\columnwidth]{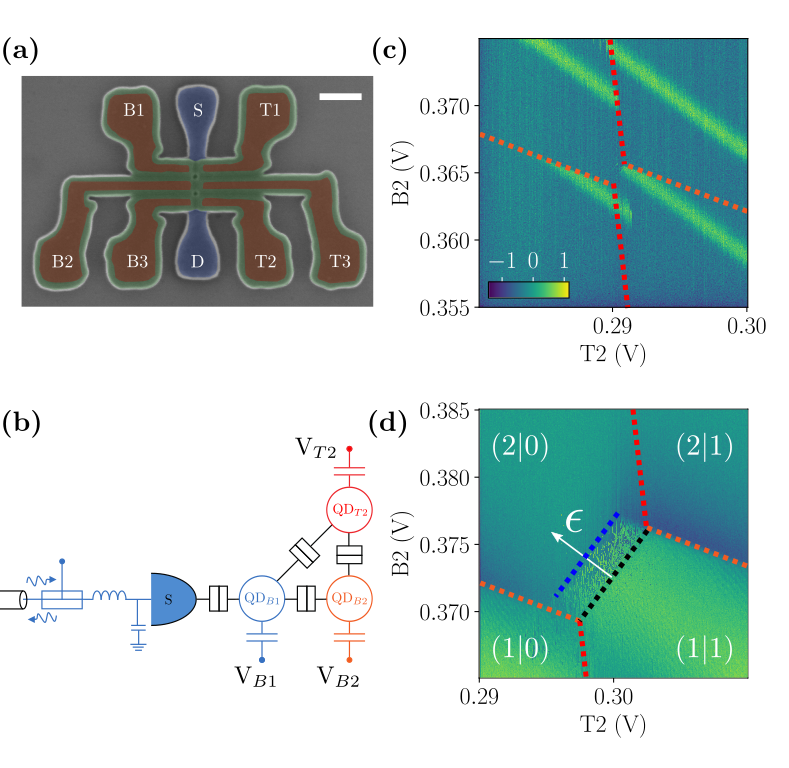}
\caption{\label{fig:Panel1}(a) False-color SEM micrograph of a device similar to the one used in this work. The quantum dot defining gates (red) are on top of the nanowire (blue) and separated by spacers (green). (b) Equivalent electrical circuit of the device in the measurement configuration used in this work. The reflectometry circuit is connected to the source. Together with quantum dot $QD_{B1}$ they form an electrometer. The sensor $QD_{B1}$ is tunnel and capacitively coupled to $QD_{B2}$ and $QD_{T2}$, allowing to sense their charge state. (c) Stability diagram of the two quantum dots $QD_{B2}$ and $QD_{T2}$. A charge transition of $QD_{B1}$ is split horizontally (vertically) by a charge transition of $QD_{B2}$ ($QD_{T2}$). Orange (red) dotted lines indicate the transition of $QD_{B2}$ ($QD_{T2}$). (d) Interdot transition of the $(2|0) - (1|1)$ regime at $\SI{300}{\milli\tesla}$. The space between the blue and black dotted lines indicate the Pauli spin blockade region, limited by the energy spacing to the next valley or orbital state.}
\end{figure}
We start tuning the sensor which is controlled by gate B1.
To minimize the tunnel coupling between the sensor and further probed dot we try to work with the smallest amount of electron possible in the dot.
However, if the number of electrons is too small the tunnel coupling between the dot and the lead is too weak to give a reflectometry signal \cite{House_2015, Gonzalez-Zalba_2015}.
We find that the optimal operation point is the degeneracy between 4 and 5 electrons in B1 (see suppl. mat. A for more detail on the tuning of the sensor dot).
Operating the sensor at this degeneracy point gives a strong capacitive coupling to nearby quantum dots which allows the detection of the first electron in the double quantum dots formed by B2 and T2\cite{Chanrion_2020, Ansaloni_2020}.
We now move to the tuning of the double quantum dot formed by B2 and T2. The gates B3 and T3 are set to $0$ to isolate the double quantum dot from the drain reservoir. To determine the charge configuration space of the double quantum dot system, we measure stability diagrams of B2-T2 and scan in a third dimension the voltage of the sensor B1 . The charge degeneracies of the sensor become visible in the B2-T2 stability diagram as broadened Coulomb peaks (see suppl. mat. A). Loading a single electron in QD$_{B2}$ leads to a splitting of the sensor signal along B2 voltage (see orange lines in Fig. \ref{fig:Panel1} (c)). The transitions of the QD defined by T2 are detected as almost vertical cuts of the sensor signal (see red line in Fig. \ref{fig:Panel1}(c)). Thus we can probe the charge occupation of QD$_{B2}$ and QD$_{T2}$ using QD$_{B1}$ and the source-reflectometry signal.

\section{Pauli Spin Blockade for Singlet-Triplet and Parity Readout}
\begin{figure}[t]
\includegraphics[width=\columnwidth]{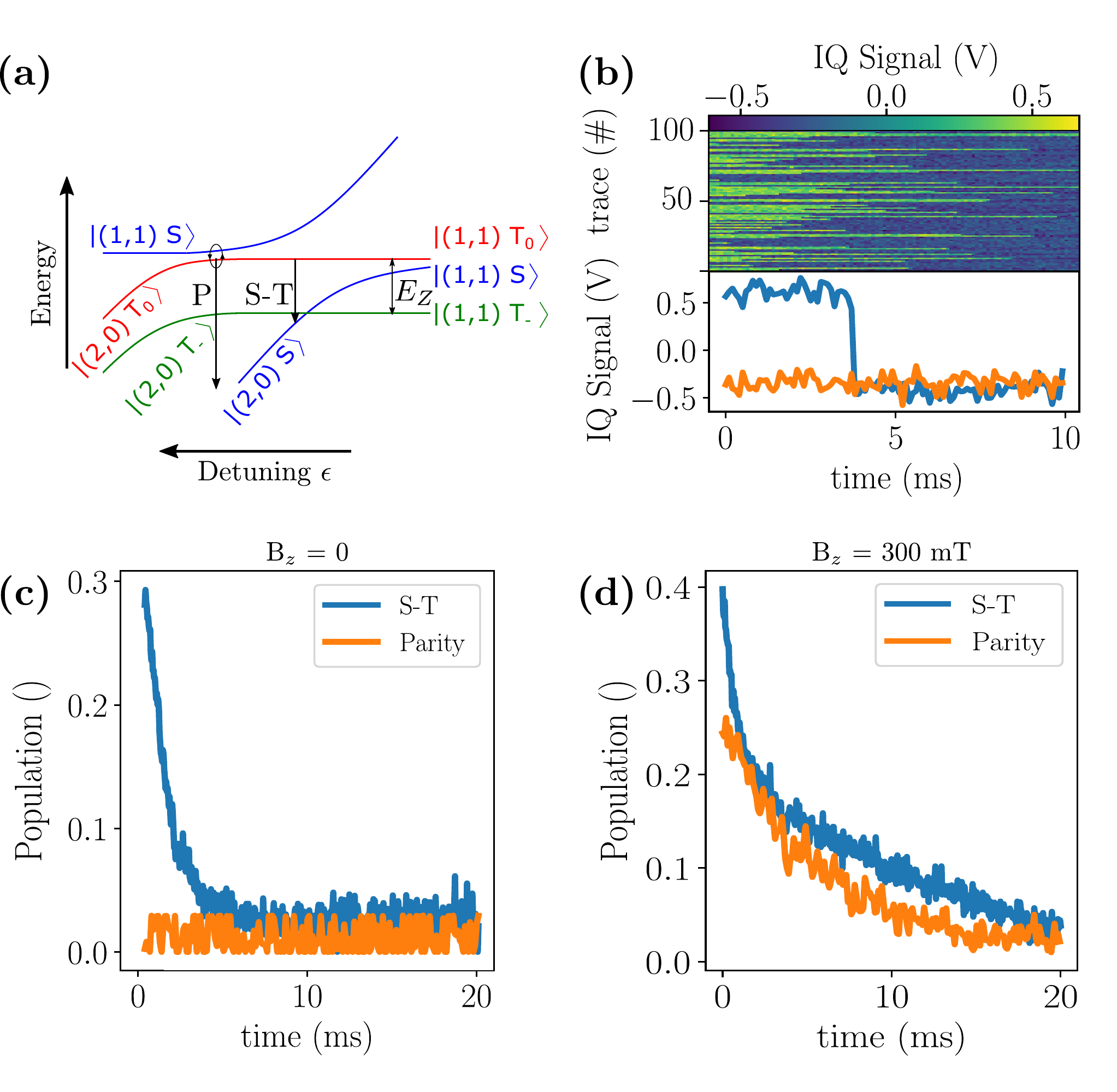}
\caption{\label{fig:Panel2}(a) Energy diagram for the lowest states of the double electron system close to the (2|0) - (1|1) transition at $B_z \neq 0$. The P labeled arrow indicates the position where the $T_0$ state relaxes fast and allows parity readout, whereas the S-T labeled arrow shows the position of S-T readout. (b) Single shot measurements of two traces at the parity readout position. A blocked state relaxing to the non-blocked state in blue and a non-blocked state in orange. (c) Comparison at zero field of the relaxation curves from the mixed triplet states to the non-blocked singlet state measured at the ST-readout (blue curve) and at the parity readout (orange curve) positions. (d) Comparison at $B_z = \SI{300}{\milli\tesla}$ of the relaxation curves from the mixed $T_-$ and $T_0$ states to the singlet state measured at the ST-readout (blue curve) and at the parity readout (orange curve) positions. The first one shows a double exponential decay with one fast decay and one slow decay, whereas the second curve just shows a single fast exponential decay.}
\end{figure}
We start by identifying the region where Pauli spin blockade can occur.
For this we tune the sensor to probe both dots simultaneously and their interdot transition corresponding to the $(2|0)-(1|1)$ charge states, depicted in \ref{fig:Panel1}(d). 
Under a magnetic field $B_z = \SI{300}{\milli\tesla}$, the $T_-$ state is the ground spin state in the (1|1) regime, whereas $S_0$ remains the ground state in the (2|0) regime as sketched in Fig. \ref{fig:Panel2}(a). 
Therefore, by scanning over the interdot transition starting from (1|1), we can identify the Pauli spin blockade region as dashed extension of the (1|1) charge state beyond the interdot transition on the (2|0) side, see \ref{fig:Panel1}(d). 
This PSB area has a finite width as the blockade is lifted when the detuning energy surpasses the valley or orbital energy separating the $T_-$ (2,0) and $T_-$ (1,1) states. After evaluation of the lever-arm to be $\approx \SI{0.05}{\electronvolt}/\SI{}{\volt}$ (see suppl. mat. F), we can estimate the valley splitting to be around $\SI{130}{\micro\electronvolt}$, in agreement with measurements performed in similar devices \cite{Spence_2022}.\\
By performing pulsed-measurement in the PSB area, we can resolve spin blockade lifting in real time  as presented in Fig. \ref{fig:Panel2}(b) . 
To investigate further the single-shot spin readout we perform two different measurements. We refer to these as singlet-triplet (S-T) readout and parity readout. The S-T readout is used to distinguish the singlet state from all triplet states. The parity readout is used to distinguish polarized spin states (or even states $T_-$ and $T_+$) from unpolarized spin states (or odd states $T_0$ and $S_0$)\cite{Yang_2020}.  
\\
We start with the standard S-T readout which allows to distinguish between the singlet and the three triplet states.
We initialise in (2|0) where the system relaxes to the ground singlet state and make a pulse to (1|1) where the singlet and triplet can mix (see suppl. mat. G).
Then, we pulse to the S-T readout position located just across the interdot transition using a non-adiabatic pulse.
At zero magnetic field and at the S-T readout position the three triplets are degenerated leading to a single exponential decay to the ground singlet state as observed on the blue curve Fig. \ref{fig:Panel2}(c) with a characteristic $T_1 = \SI{0.9}{\milli\second}$.
At finite magnetic field, we initialise in (2|0) where the system relaxes to the ground singlet state and ramp to (1|1). Using a Landau-Zener experiment at the $S-T_-$ anticrossing (see suppl. mat. E) combined with $S-T_0$ mixing (see suppl. mat. G), we set the ramp sweep rate to obtain an initial state which contains a similar fraction of $T_-$ and $T_0$ with $S_0$. We obtain the blue curves on Fig. \ref{fig:Panel2}(d) for $B_Z = \SI{0.3}{\tesla}$.
At finite magnetic field the $T_-$ and $T_0$ are split in energy leading to different relaxation characteristic times as shown by the double exponential decay on the blue curve with characteristic times $T_1 = \SI{0.9}{\milli\second}$ and $T_1 = \SI{32}{\milli\second}$.
These signatures show that it is possible to distinguish between $S_0$ and the triplet states. 
To obtain further information on the different triplet populations we can rely on their different relaxation dynamics and readout at different timescales.
For instance probing the spin state at short time allows to distinguish between singlet and triplet and probing $\SI{3}{\milli\second}$ later when $T_0$ has relaxed allows to get information on the remaining $T_-$ population.
However, such measurement leads to poor fidelity in $T_-$ readout, $\approx\SI{95}{\percent}$, due to the small contrast between $T_0 \rightarrow S_0$ and $T_- \rightarrow S_0$ relaxation rates(see suppl. mat. D).
To improve this fidelity we propose to move to a second measurement point where the $T_0$ relaxation rate is drastically increased to perform a parity readout. 
\\
We perform the same initialisation for the parity readout. The measurement position is now further in the (2|0) regime, where the $T_0$ state relaxes much faster. 
In contrast to the S-T readout, at zero magnetic field, we cannot observe any blocked state leading to a flat relaxation curve at the measurement point, see the orange curve in Fig. \ref{fig:Panel2}(c). At finite magnetic field, see the orange curve in Fig. \ref{fig:Panel2}(d), similarly, the rapid exponential decay has disappeared, leaving only a slow relaxation attributed to $T_-$ population.
In both cases, there is no signature of a $T_0$ relaxation which leads us to the conclusion that the $T_0$ has relaxed prior to any measurement due to mixing with the excited $S_0$ state followed by charge relaxation\cite{Seedhouse_2021}.

\section{Fidelity benchmarking of parity readout}
\begin{figure}[t]
\includegraphics[width=\columnwidth]{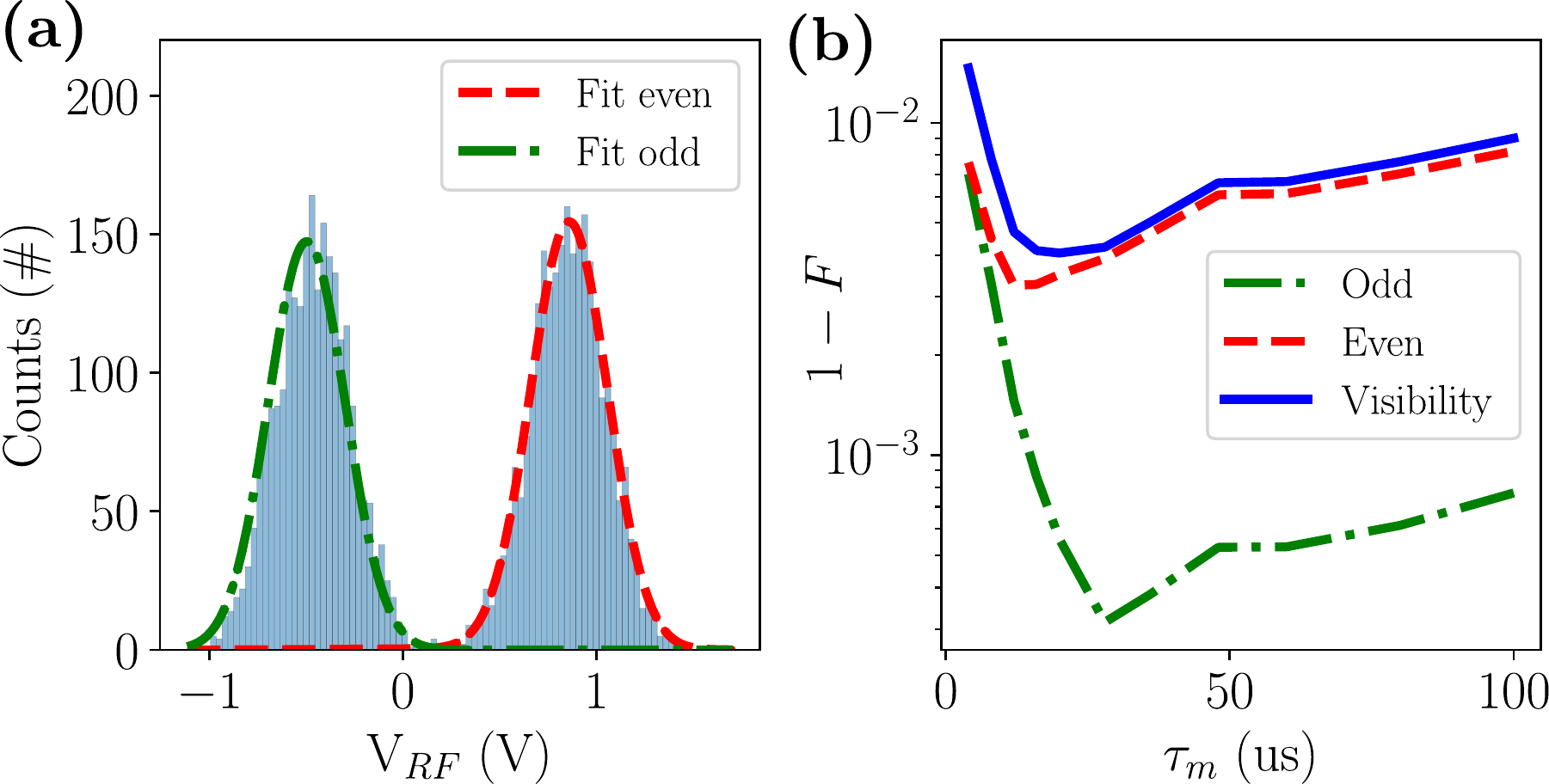}
\caption{\label{fig:Panel3}(a) Histogram of 10000 parity readout measurements at a readout speed of $\SI{50}{\kilo\hertz}$. The state was initialized with $\approx \SI{50}{\percent}$ singlet (triplet) probability. The green (red) curve are fitting the odd (even) state population, following Barthel et al. \cite{Barthel_2009}.(b) Fidelity/Visibility error as a function of integration time $\tau _m$. The optimal fidelities/visibility is found for $\tau_m = \SI{20}{\micro\second}$, due to noise broadening error for faster integration and relaxation error for slower integration.}

\end{figure}
In the following section, we want to discuss the optimization of our parity readout, which could as well be used to optimize the ST-readout. The readout time is constrained by the relaxation time $T_1$ at the measurement position. Another parameter to optimize is the RF-power of the readout, affecting the back action on the double dot, that can drive relaxation\cite{Reilly_2007}. We benchmark our readout fidelity by varying the RF power as well as the integration time. Performing experiments where we prepare approximately $\SI{50}{\percent}$ even/odd state ratio by relaxation to the respective ground state, we measure the signal distribution for 1000 repetitions. We follow Barthel et al. \cite{Barthel_2009} to calculate the fidelities by fitting a normal distribution to the signal distribution of the odd state and a normal distribution with a decay term to the signal distribution of the even state (accounting for relaxation). Plotting the SNR as a function of power and integration time (see suppl. mat. C), we find an optimal power at around $\SI{-91}{dBm}$. We accumulate 10000 single shot traces for each integration time $\tau_m$ to ensure sufficient sampling of the signal distribution. A histogram of the signal distribution for $\tau_m = \SI{20}{\micro\second}$ is shown in FIG. \ref{fig:Panel3}(a). FIG. \ref{fig:Panel3} (b) depicts a plot of the fidelities and visibility as a function of integration time $\tau_m$. We find an optimal integration time of $\tau_{m,opt} = \SI{20}{\micro\second}$ with an odd (even) fidelity of $\SI{99.98}{\percent}$ ($\SI{99.83}{\percent}$). This results in a visibility of $\SI{99.79}{\percent}$. Reducing the integration time to $\SI{4}{\micro\second}$, which is the limit of our measurement bandwidth, the fidelities are slightly lower with $\SI{99.57}{\percent}$ ($\SI{99.56}{\percent}$) for odd (even) and a visibility of $\SI{99.13}{\percent}$, still being above $\SI{99}{\percent}$. The lower fidelity at short integration times is in good agreement with the expected noise broadening, decreasing the SNR by $\propto \sqrt{\tau_m}$. The exponential decrease of fidelities, found for longer integration times, agrees as well with the expected exponential growth of leaked even state into odd state due to relaxation.  

\section{State preparation and error analysis}
\begin{figure}[t]
\includegraphics[width=\columnwidth]{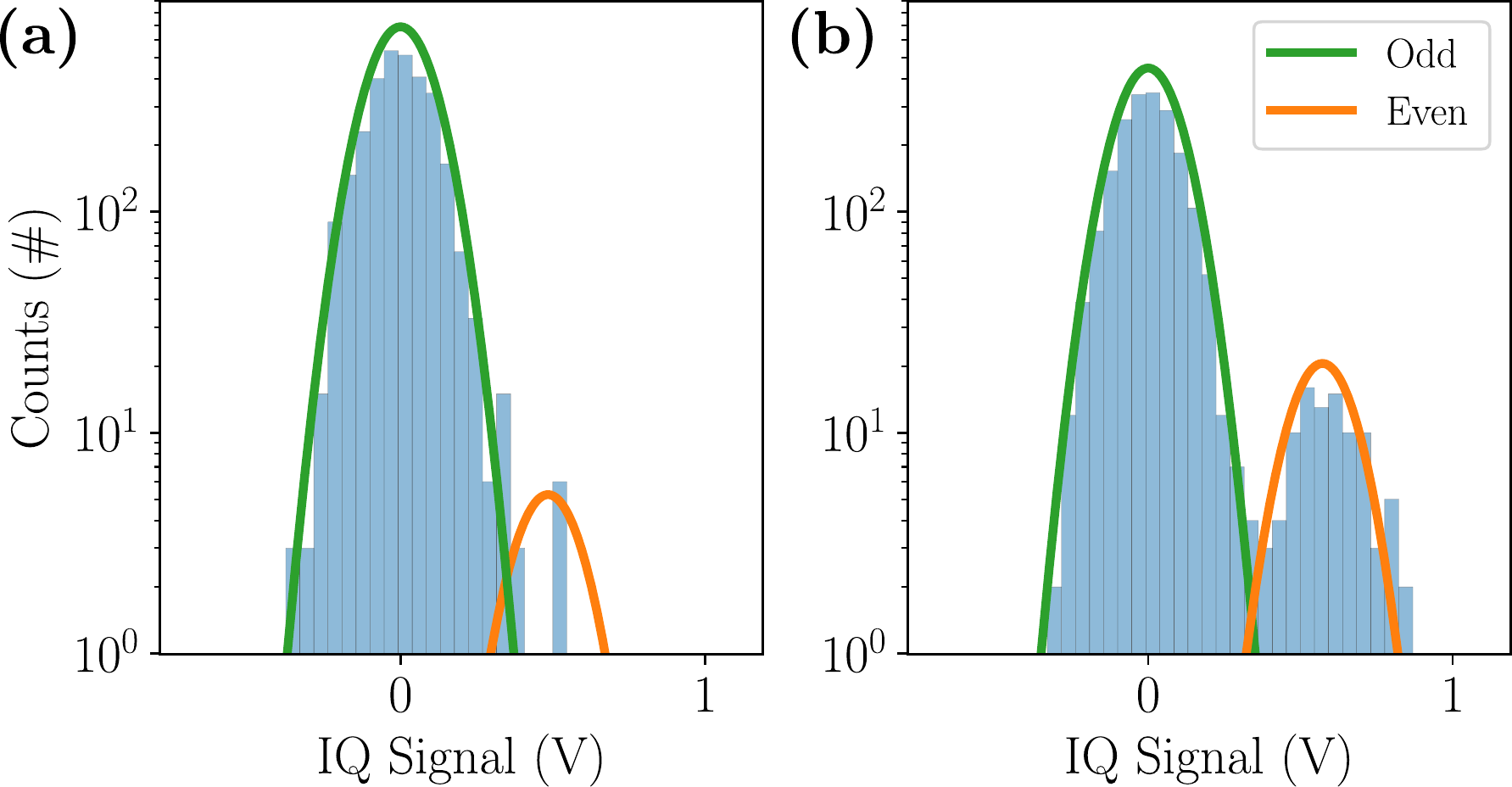}
\caption{\label{fig:Panel4}(a) Histogram of $S_0$ initialization.(b) Histogram of an initialized $S_0$, followed by a transfer to (1|1) with a direct return to the measurement position.}

\end{figure}
The total fidelity of state preparation and measurement (SPAM) sums up two DiVincenco criteria. We break down the error contributions into three different types: initialization, transfer to the regime of single qubit operation (1|1) and readout. The latter has been characterized in the previous section.
To identify the initialization error, we start by initializing in (1|0) and load a second electron into QD$_{B2}$ by pulsing into (2|0), where we allow relaxation to the ground state by waiting for $\SI{10}{\milli\second}$. 
Then we pulse after this initialization phase to the readout measurement position and measure the spin state. We find the signal distribution depicted in FIG. \ref{fig:Panel4}(a), indicating a $S_0$ initialization fidelity of $\SI{99.6}{\percent}$.\\ 
We now investigate errors due to transfer in the (1|1) where the two electrons can be decoupled. 
We initialize a singlet state in (0|2) as previously described. We then transfer one electron by performing first a non-adiabatic pulse to avoid the $S-T_-$ anti-crossing (see suppl. mat. E) followed by an adiabatic ramp deep in (1|1) to avoid mixing S with $T_0$. Without waiting, we pulse from this position to the measurement position. Considering the time per instruction in our sequence, the total time spent in (1|1) is $\approx \SI{20}{\micro\second}$, negligible with regards to the relaxation time $T_1 > \SI{1}{\second}$ in (1|1) (see suppl. mat. D). We find a $S_0$ population of $\approx\SI{95.6}{\percent}$ for the transfer measurement by using the signal distribution depicted in \ref{fig:Panel4}(b). We attribute the $\approx \SI{4}{\percent}$ difference between these two experiments to leakage during the transfer through the $S_0$-$T_-$ anti-crossing.
This assumption is supported by a Landau-Zener type of experiment where the transfer from (0|2) to (1|1) is performed at different rate (see suppl. mat. E). The presence of the anti-crossing leads to a sweep-rate-dependent return singlet probability.

\section{Conclusion}
We have shown how we can operate a triple quantum dot to perform high fidelity single shot readout of a double dot system. Thanks to a strong capacitive coupling and reflectometry method we have been able to achieve spin readout fidelity above $\SI{99.9}{\percent}$ ($\SI{99}{\percent}$) in $\SI{20}{\micro\second}$ ($\SI{4}{\micro\second}$). Using the spin readout we have characterized the different error sources during initialization and displacement of electrons between the (2|0) and (1|1). 
Finally, by adjusting the measurement position in detuning we can alternatively use singlet-triplet or parity readout. Combining sequentially these two readouts can be of strong interest in order to extract the full spin information of a 2-qubit system. As proposed by \cite{Nurizzo_2022}, to achieve such complete readout, we could start with a S-T readout to distinguish singlet from all triplets. Followed by a parity readout, it would allow to distinguish the unpolarized triplet ($T_0$) from the two polarized ones($T_-$ and $T_+$). Finally, an adiabatic transfer which swaps the $T_-$ and $S_0$ population followed by a ST or parity readout will allow to differentiate between the two polarized triplets.\\
\textit{Note added}. During the preparation of this manuscript, we became aware of a recent experimental observation of a PSB in a similar device \cite{Oakes_2022}.

\section*{Acknowledgment}
We acknowledge support for the cryogenic apparatus from W. Wernsdorfer, E. Bonet and E. Eyraud. We acknowledge technical support from L. Hutin, D. Lepoittevin, I. Pheng, T. Crozes, L. Del Rey, D. Dufeu, J. Jarreau, C. Hoarau and C. Guttin. 
D.J.N. acknowledges the GreQuE doctoral programs (grant agreement No.754303). The device fabrication is funded through the Mosquito project (Grant agreement No.688539).
This work is supported by the Agence Nationale de la Recherche through the CRYMCO project and the CMOSQSPIN project. This project receives as well funding from the project QuCube(Grant agreement No.810504) and the project QLSI (Grant agreement No.951852).
\bibliography{bibliography}
\newpage

\section{Suppl. Mat}
\subsection{Charge state of double quantum dot system}
\begin{figure*}[t]
	\includegraphics[width=\textwidth]{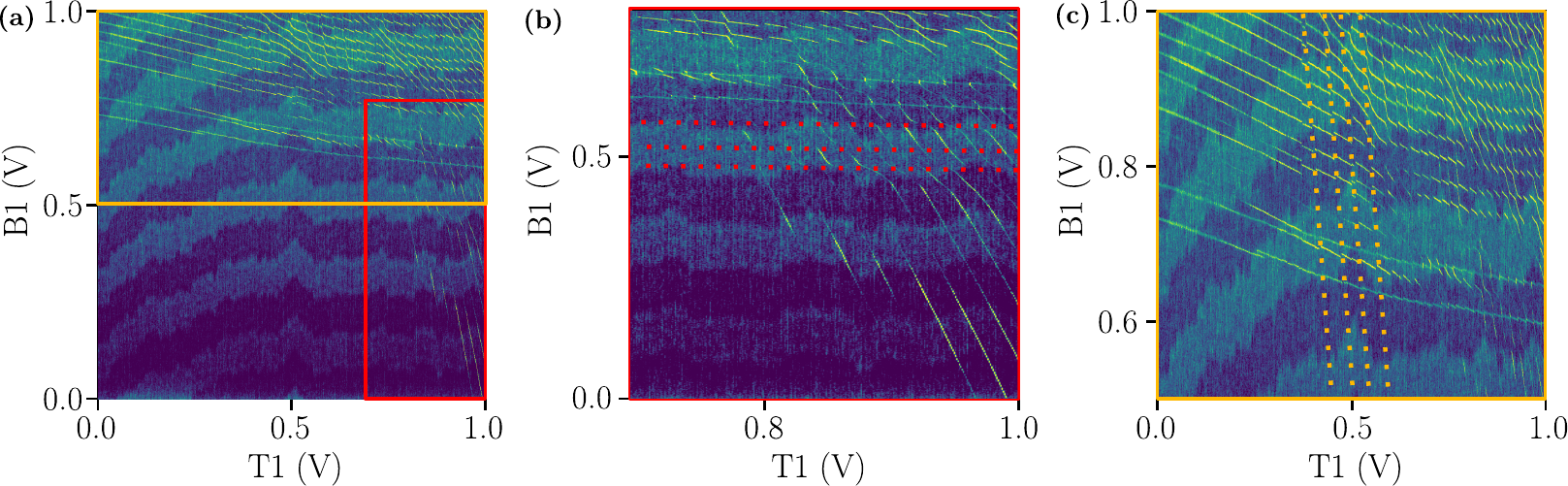}
	\caption{\label{fig:panel1}(a) Charge stability diagram of the two sensing dots QD$_{B1}$ and QD$_{T1}$. (b) Zoom of the region in the red rectangle in (a) with indicated transitions of QD$_{B1}$, sensed by QD$_{T1}$. (c) Zoom of the region in the orange rectangle in (a) with indicated transitions of QD$_{T1}$, sensed by QD$_{B1}$.}
\end{figure*}
\label{sec:suppl_charge_state}
\begin{figure*}[t]
	\includegraphics[width=\textwidth]{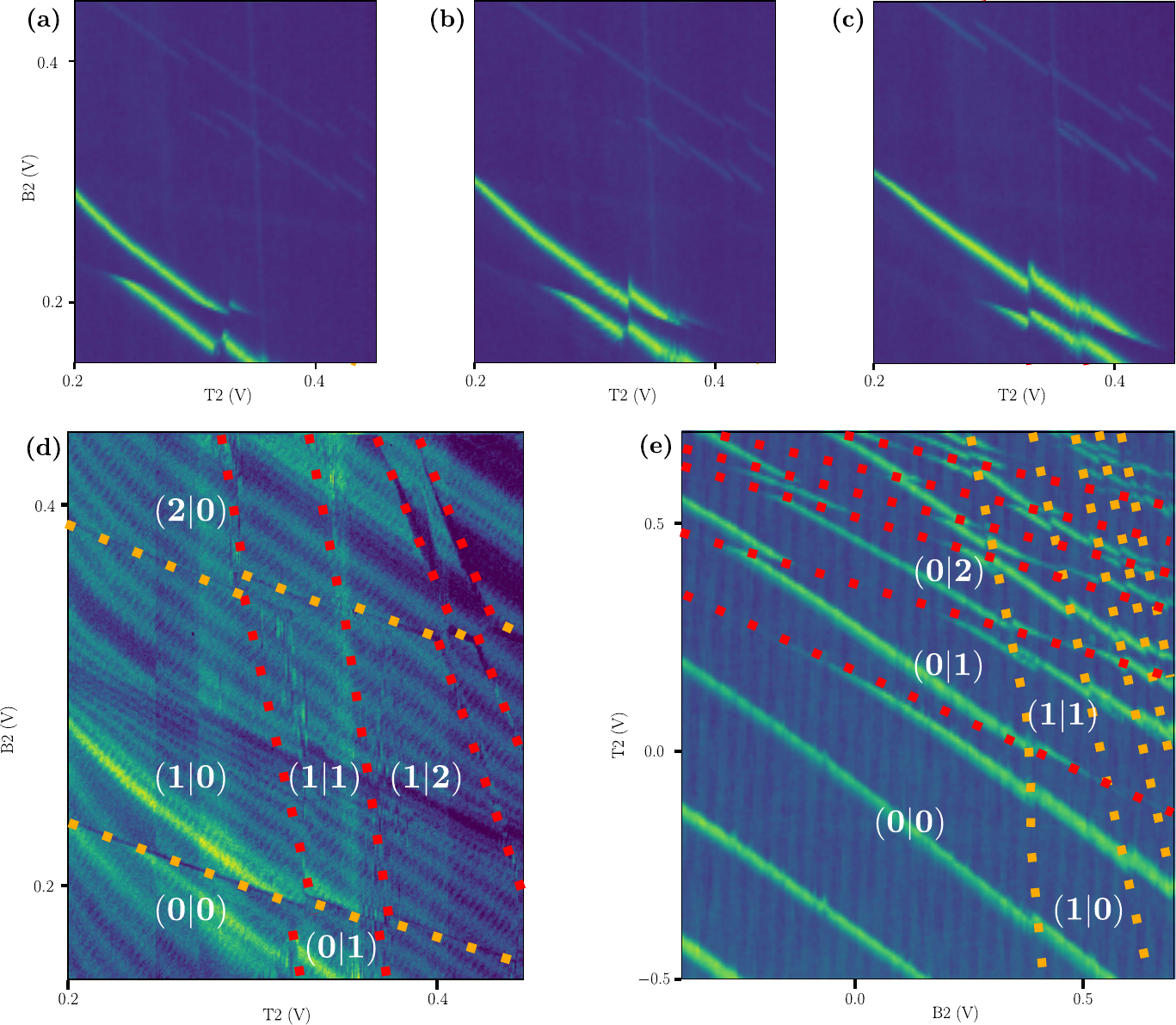}
	\caption{\label{fig:panel2}(a) Stability diagram of the two gates B2 and T2 with the sensor B1 set to $\SI{0.662}{\volt}$. A single charge degeneracy line of the sensor is visible. When a charge transition of one of the two QDs B2 or T2 is aligned with the sensor degeneracy point, the sensor line shows a sharp horizontal (vertical) discontinuity for a transition of B2 (T2). (b) and (c) depict the same stability diagram with the gate voltage on the sensor gate set to $\SI{0.6615}{\volt}$ and $\SI{0.661}{\volt}$ respectively. (d) Overlap of 20 stability diagrams with the sensor gate voltage ranging from $\SI{0.67}{\volt}$ - $\SI{0.657}{\volt}$. (e) Stability diagram using QD$_{T1}$ as sensor. Using multiple sensor degeneracy points in a single stability diagram allows the identification of the charge regimes of the double quantum dot system.}
\end{figure*}
\begin{figure}[t]
	\centering
	\includegraphics[width=\columnwidth]{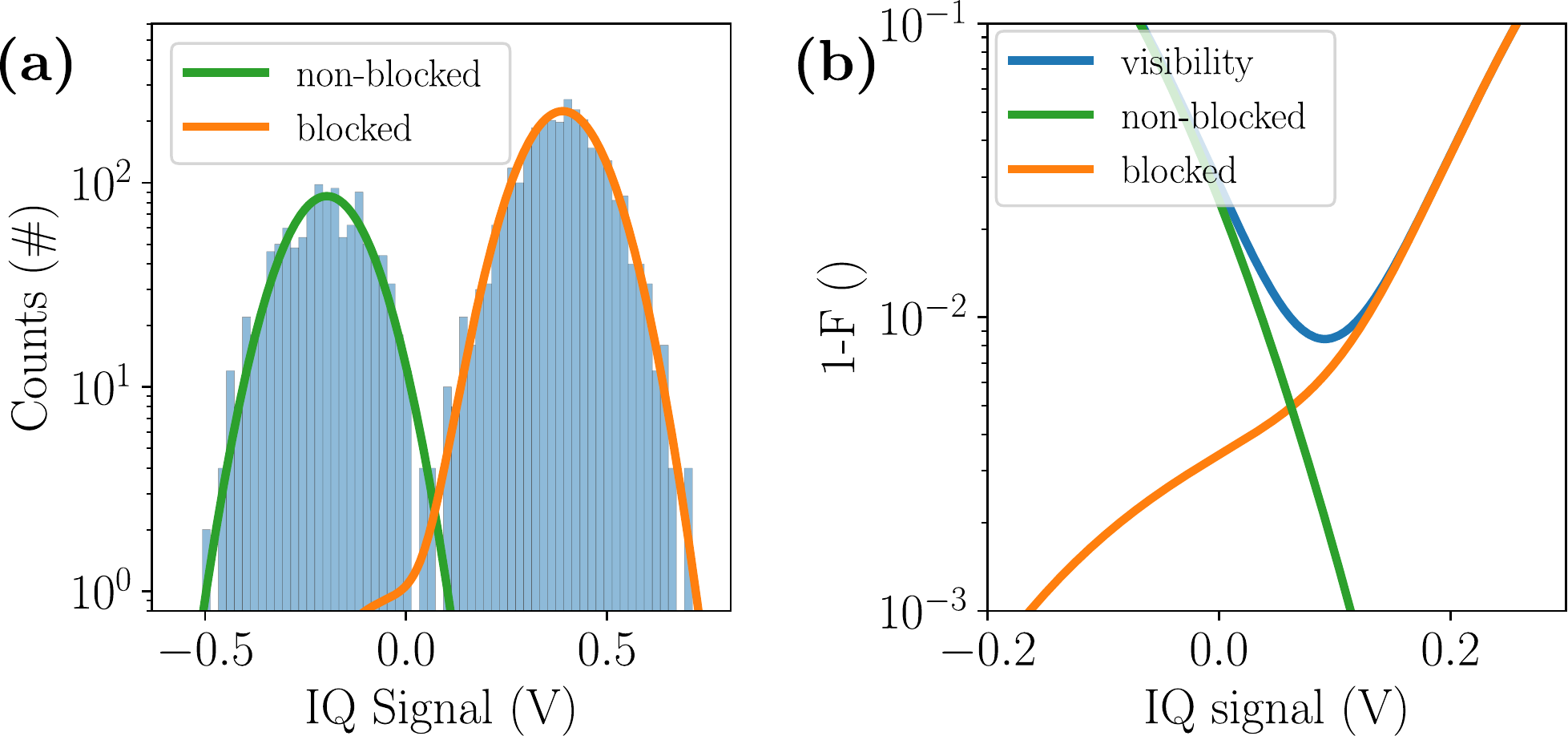}
	\caption{\label{fig:panel3}(a) Histogram of 10000 single shot traces at the parity readout position. Fitted distribution of blocked (orange) and non-blocked (green) state. (b) Fidelity and visibility error defined as 1-F for the fits in (a).}
\end{figure}
We measure a stability diagram of the two potential sensing dots to decide on which one we use as a sensing dot. The stability diagram is depicted in FIG. \ref{fig:panel1} (a) with a zoom in the respective region of interest in (b) and (c) for QD$_{B1}$ and QD$_{T1}$.
We use QD$_{B1}$ as a sensing dot as it shows a strong signal at a low number of electrons. 
Using the charge degeneracy points of the sensing dot as a charge sensor for the two center quantum dots allows to determine the number of charges in the quantum dots. FIG. \ref{fig:panel2} depicts three stability diagrams for different sensor voltages. The transitions of QD$_{B2}$ (QD$_{T2}$) are indicated as orange (red) lines in. FIG. \ref{fig:panel2}(d) allows to map out all transitions of the sensed dots by overlaying 20 stability diagram. Using multiple degeneracy points of the sensor allows to identify the different charge regimes in a single stability diagram as depicted in FIG. \ref{fig:panel2}(e) (here T1 is used as sensor). In the main text, we use the first degeneracy point of the QD defined by B1 as sensor. In this configuration, the sensor is weakly tunnel coupled to the QD of T2, reducing lifting of PSB by co-tunneling through the sensor dot.
\subsection{Fidelity/Visibility definitions}
Following the analysis of Barthel et al. \cite{Barthel_2009}, we fit the signal distribution of the PSB measurement using:
\begin{align}
	n_S(V_{rf}) &= \frac{1-\expval{P_T}}{\sqrt{2\pi}\sigma}e^{-\frac{V_{rf}-V^S_{rf}}{2\sigma ^2}} ,\label{equ:signal_dist_1}\\
	n_T(V_{rf}) &= \frac{\expval{P_T}}{\sqrt{2\pi\sigma}}e^{-\frac{\tau_m}{T_1}}e^{-\frac{(V_{rf}-V^T_{rf})^2}{2\sigma ^2}} \nonumber \\
	&+ \int _{V_{rf}^S} ^{V_{rf}^T} \frac{\tau _m}{T_1}\frac{\expval{P_T}}{\Delta V_{rf}}e^{-\frac{V-V_{rf}^S}{\Delta V_{rf}}\frac{\tau _m}{T_1}}e^{-\frac{(V_{rf}-V)^2}{2\sigma ^2}} \frac{dV}{\sqrt{2\pi}\sigma},\label{equ:signal_dist_2}
\end{align}
where $\expval{P_T}$ is the triplet probability, $V^S_{rf}$ ($V^T_{rf}$) is the signal expectation value for the non-blocked (blocked) state, $\sigma$ is the standard deviation of the Gaussian signal distribution, $\tau _m$ is the measurement integration time, and $T_1$ is the lifetime at the measurement position. While equation \ref{equ:signal_dist_1} is a Gaussian distribution, describing the non-blocked state signal distribution, the excited state is given by equation \ref{equ:signal_dist_2}, the convolution of a Gaussian distribution with an exponential decay. We use these functions to fit the signal distribution of our PSB measurements. An example is given in FIG. \ref{fig:panel3} (a).\\
The definition of fidelities allows a simple metric to estimate the error of the signal assignment as blocked (non-blocked). The singlet and triplet fidelities are defined as
\begin{align}
	F_S &= 1 - \int _{V_T}^\infty n_S (V)dV, \label{equ:fidelity_S}\\
	F_T &= 1 - \int _{-\infty}^{V_T} n_T (V)dV, \label{equ:fidelity_T},
\end{align}
where $V_T$ is the threshold that separated the identification as blocked/non-blocked state. The visibility is then defined as 
\begin{align}
	V = F_S + F_T -1.
\end{align}
To optimize the threshold, one calculates the maximum of the visibility. FIG. \ref{fig:panel3}(b) depicts the error of the fidelities and visibility for the fits from (a) close to the maximal visibility.

\subsection{Reflectometry backaction}
\label{sec:suppl_backaction}
\begin{figure*}[t]
	\includegraphics[width=\textwidth]{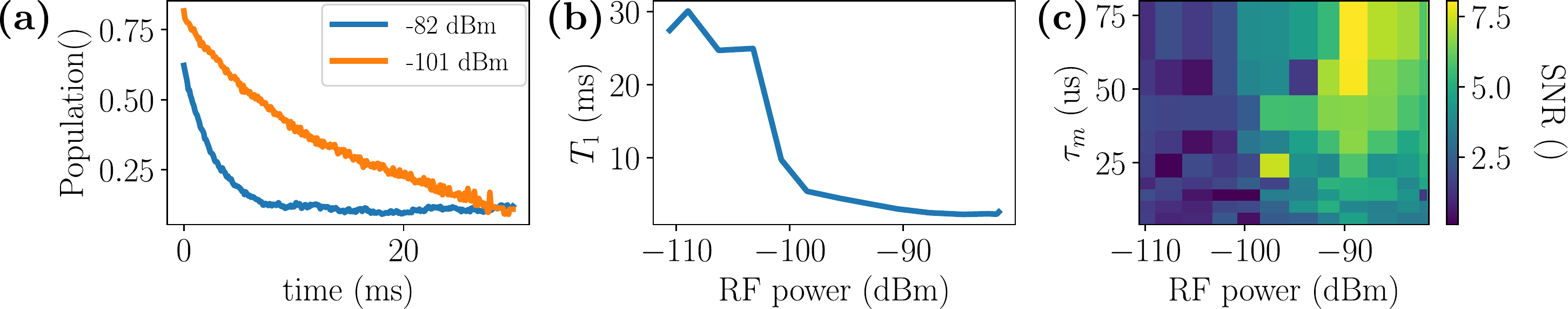}
	\caption{\label{fig:panel4}(a) Relaxation of the even state at the parity readout position for $\SI{-71}{dBm}$ and $\SI{-91}{dBm}$ RF power. The relaxation is faster for higher RF-power. (b) Lifetime $T_1$ of even states as a function of RF input power at the parity readout position. The graph shows a sharp decrease in $T_1$ of one order of magnitude around $\SI{-100}{dBm}$. (c) SNR as a function of integration time $\tau _m$ and RF power.}
\end{figure*}
We investigate the backaction of the sensing mechanism on the spin state by performing fidelity measurements as a function of the RF-power. We find that for $> \SI{-100}{dBm}$, the lifetime at the measurement position is strongly decreased by one order of magnitude (see FIG. \ref{fig:panel4} (a) and (b)). However, increasing the RF-power goes along with an increase in signal strength. We map the SNR as a function of RF-power and integration time in FIG. \ref{fig:panel4} (c). The best SNR is found at around $\SI{-90}{dBm}$, where the lifetime $T_1$ at the measurement position is $\approx \SI{3}{\milli\second}$. 
\subsection{Lifetime of $S_0$ in (1|1)}
\label{sec:suppl_lifetime}
\begin{figure}[t]
	\includegraphics[width=\columnwidth]{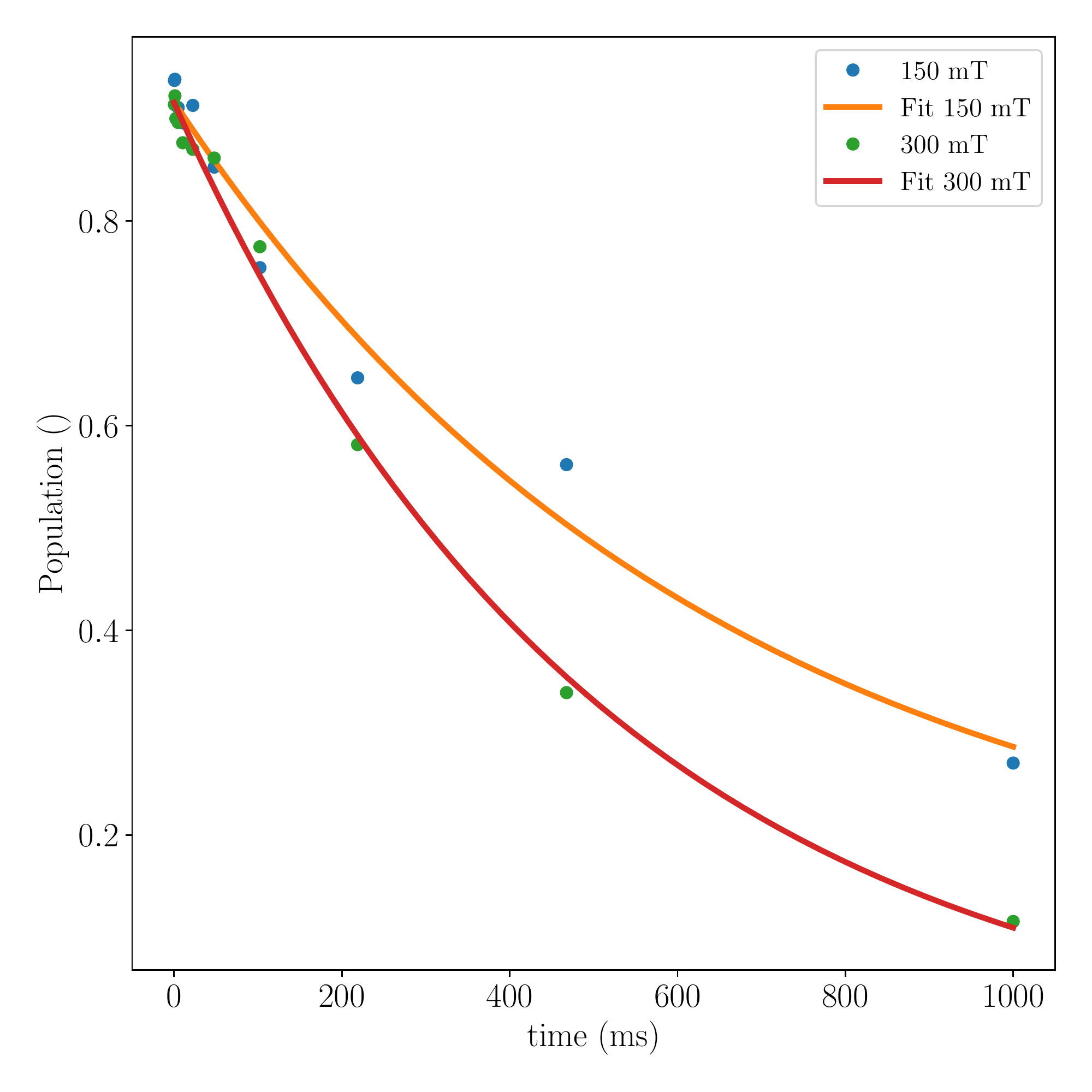}
	\caption{\label{fig:Panel5_suppl}Singlet population as a function of waiting time in (1|1) for a magnetic field of $\SI{150}{\milli\tesla}$ (blue dots) and $\SI{300}{\milli\tesla}$ (green dots). Fitted $S_0$ relaxation for $\SI{150}{\milli\tesla}$ (orange) ($\SI{300}{\milli\tesla}$ (red)) giving a lifetime $T_1 \approx \SI{640}{\milli\second}$ ($\SI{520}{\milli\second}$).}
\end{figure}
While we initialize in the $S_0$ state, spin operations take place in the (1|1) regime where $T_-$ is the ground state. Therefore, the relaxation of $S_0$ in (1|1) must be much slower than the spin manipulation. We measure the $S_0$ relaxation by preparing a $S_0$ state, followed by pulsing in the (1|1) regime. Deep in the (1|1) regime where we can assume that the quantum dots are completely decoupled, we wait for a given time $\tau$ ranging from $\SI{0.1}{\milli\second}$ to $\SI{3}{\second}$. After, we pulse to the parity readout position and measure the spin state. We fit the resulting signal distribution from 2000 data traces. The $T_- (T_+)$ population from these measurements is depicted in FIG. \ref{fig:Panel5_suppl} . We fit an exponential function with a decay time $T_1 \approx \SI{1.6}{\second}$. This relaxation time is much longer than typical times of operation in (1|1) which are typically not longer than a few $\si{\micro\second}$, around six orders of magnitude shorter than the relaxation time. The high temperature of operation compared to the magnetic field of $\SI{150}{\milli\tesla}$ leads to a relatively high residual $S_0$ population in this experiment.

\subsection{Landau-Zener experiment}
\label{sec:suppl_LZ}
\begin{figure}[t]
	\includegraphics[width=\columnwidth]{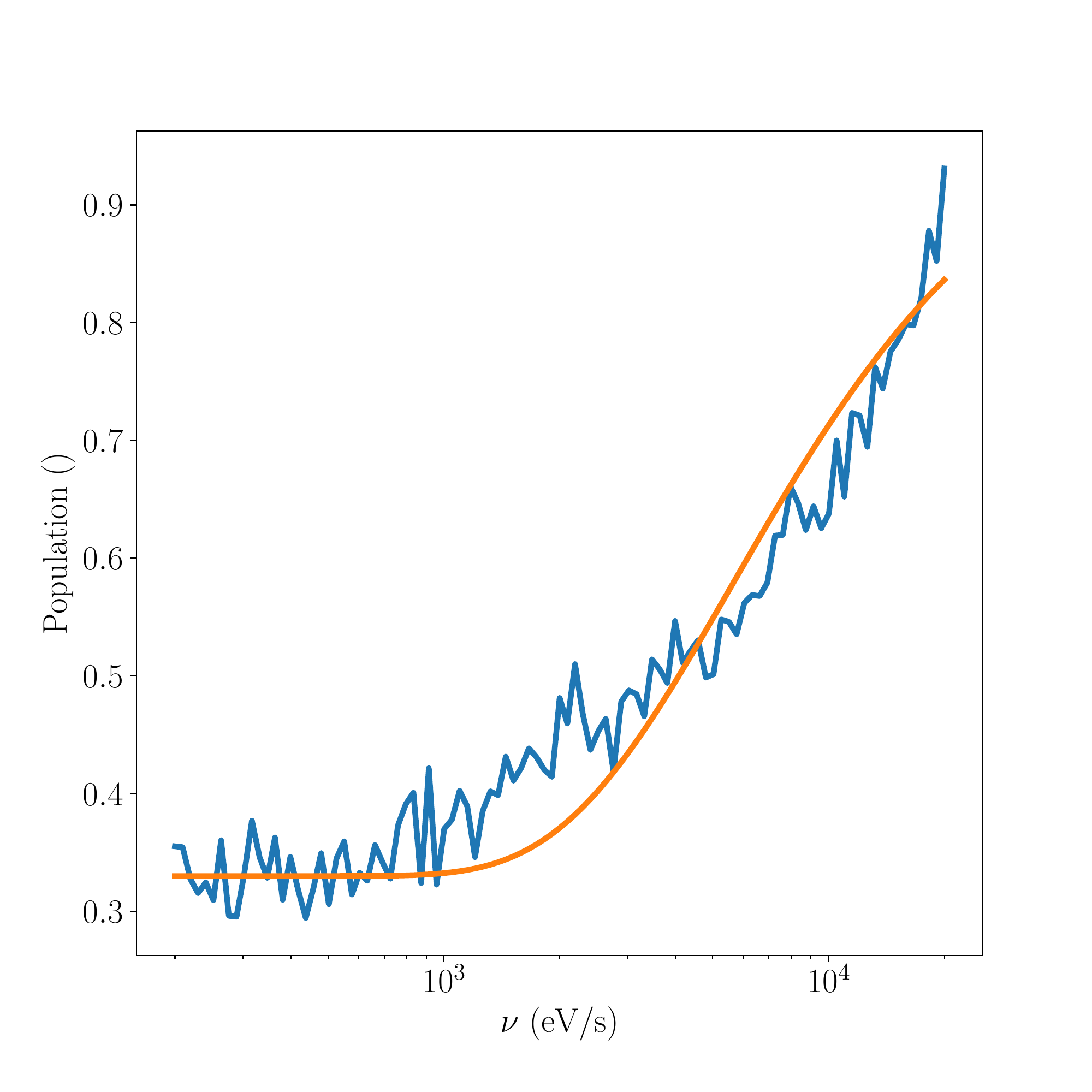}
	\caption{\label{fig:lz_pop_suppl}Singlet population as a function of transfer speed. The inset shows a schematic pulse sequence of the Landau-Zener experiment. While we keep the amplitude $\Delta \epsilon$ constant, we vary the ramp time $\tau$. The transfer speed is than calculated from $\nu = \frac{\alpha e \Delta \epsilon}{\tau}$, with $\alpha$ and $e$ the gate lever arm and elementary charge, respectively.}
\end{figure}
The transfer from the (2|0) regime to the (1|1) involves the passage of the $S_0$ - $T_-$ anti-crossing for $B_z \neq 0$. We perform a Landau-Zener experiment to estimate the fraction of non-adiabatic transfer. We initialize in $S_0$ and ramp with amplitude $\Delta \epsilon$ from (2|0) to (1|1) and return non-adiabatically to the measurement position. The pulse schematic is depicted in the inset in FIG. \ref{fig:lz_pop_suppl} . We perform this experiment with different ramp speeds and calculate the transfer speed as $\nu = \frac{\alpha e \Delta \epsilon}{\tau}$, with $\alpha$ and $e$ the gate lever arm and elementary charge, respectively. The experiment was performed at a base temperature of $\SI{100}{\milli\kelvin}$ and a magnetic field $B_z = \SI{300}{\milli\tesla}$. We find the expected monotonous increase of singlet conservation with higher transfer speed and extract a $S_0$ - $T_-$ avoided crossing of $\SI{120}{\mega\hertz}$. For very slow transfer, the population tends towards more and more population of the triplet ground state. The residual $S_0$ population could arise from charge noise which induces rapid fluctuations in the vicinity of the anticrossing reducing the maximum transfer probability \cite{Nichol2015}.
\subsection{Measuring gate lever arm}
\begin{figure}[t]
	\includegraphics[width=\columnwidth]{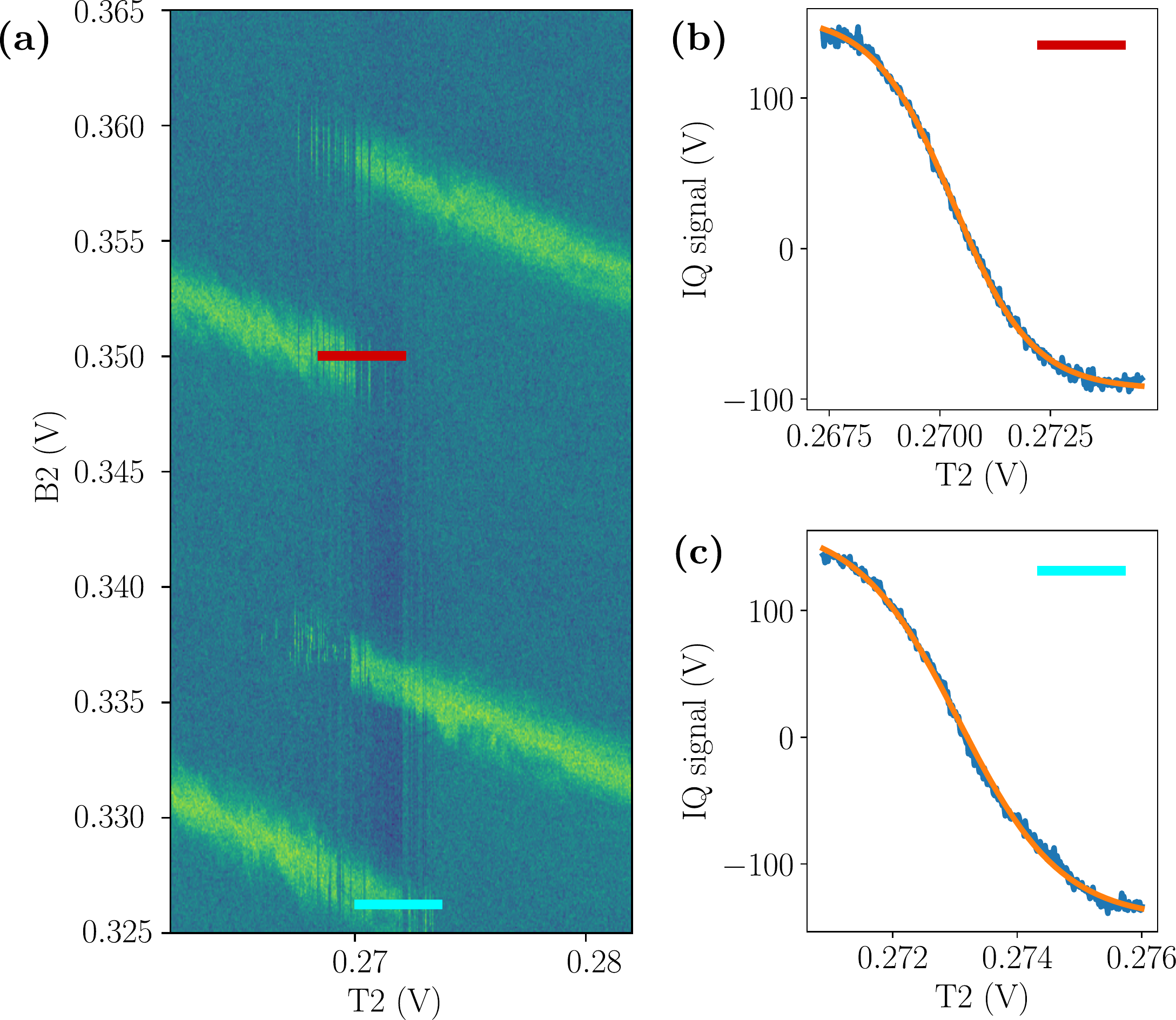}
	\caption{\label{fig:alpha_factor}(a) Stability diagram showing a charge transition of T2. (b) and (c) depict traces along the charge degeneracy point of QD$_{T2}$ and a respective fit (orange) to the sensor signal. The depicted traces in (b) and (c) are averaged over 500 measurements.}
	
\end{figure}
We determine the lever arm using the method proposed by Rossi et al. \cite{Rossi_2012}. We average traces along the charge transition of T2 indicated in FIG. \ref{fig:alpha_factor} (a). The conversion between gate voltage and dot potential is given by 
\begin{align}
	E-E_0 = \alpha e(V_{G}-V_0),    
\end{align}
where $E$ is the energy, $E_0$ is the energy of the system at the charge degeneracy point, $\alpha$ is the gate lever arm, $e$ is the elementary charge, and $V_G$ is the gate voltage with respect to the voltage of the degeneracy point of the charge transition $V_0$.
For this transition the condition $E_C \gg k_B T_e \gtrsim \Delta \epsilon$, with $E_C$ the charging energy, $k_B$ the Boltzmann constant, $T_e$ the electron temperature and $\Delta \epsilon$ the single-particle level separation, is fulfilled. We can thus approximate the transition broadening using the Fermi-Dirac distribution,
\begin{align}
	f(E-E_0) &= \frac{1}{1+e^{-(E-E_0)/k_B T_e}} \nonumber \\
	&= \frac{1}{1+e^{\alpha e(V_G - V_0)/k_B T_e}} .
\end{align}
We extract from the charge transition fits in FIG. \ref{fig:alpha_factor} (b) and (c) a gate lever arm of $\alpha = 0.05 \pm 0.002$.

\subsection{S-$T_0$ mixing}
\label{sec:mixing}
\begin{figure}
	\includegraphics[width=\columnwidth]{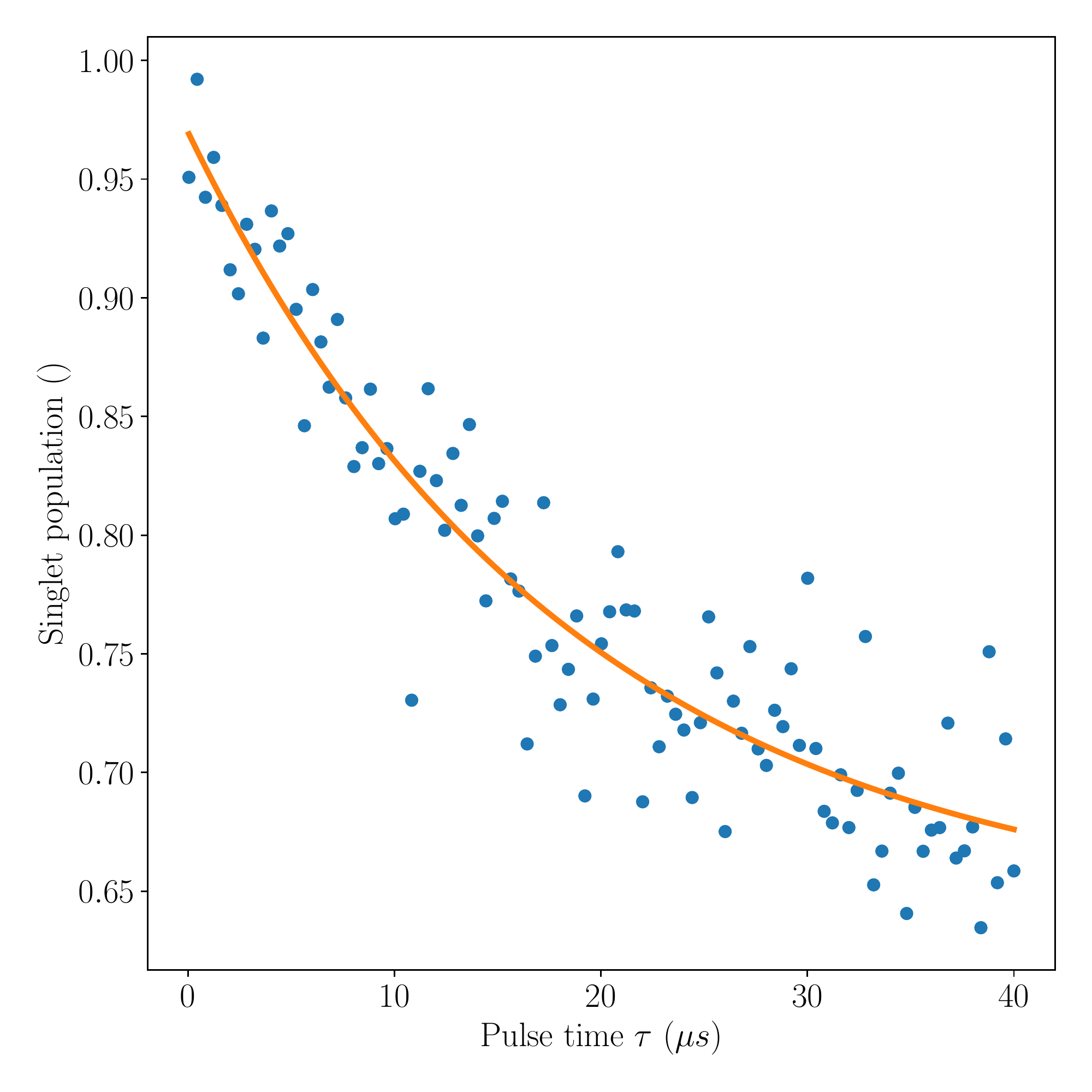}
	\caption{\label{fig:suppl_mixing}Singlet population as a function of pulse duration measured at the S-T readout position. The square pulse is applied to gate B2 to move the electrons from the (0|2) to the (1|1) regime. We interpret the decay as a mixing between the S and $T_0$ states with a characteristic time, $18.5 \pm \SI{2.5}{\micro\second}$, in agreement with the spin decoherence in isotopically purified $^{28}Si$ measured in \cite{Harvey_2019}.}
\end{figure}
We perform an exchange experiment similar to \cite{Maune_2012, Harvey_2019} by preparing a singlet state through relaxation to the ground state in (0|2). After we pulse the T2 and B2 gates to the measurement position of ST-readout, followed by a square pulse on the B2 gate to move electrons deep in (1|1). After the AWG pulse, we measure the spin state using Pauli spin blockade at the S-T readout. We plot the singlet population as a function of pulse duration in FIG. \ref{fig:suppl_mixing}. The singlet population as a function of pulse duration can be fitted with an exponential decay, indicating a characteristic time scale of $18.5 \pm \SI{2.5}{\micro\second}$. The exponential decay and the convergence towards a population of $\SI{50}{\percent}$ is an indication of the $S_0$(1|1) and $T_0$(1|1) occurs. However, no spin-orbit induced coherent oscillations are observed. This could be explained by a quenched difference of g factor when the magnetic field is applied perpendicular to the nanowire axis as found in planar MOS silicon double quantum dot with perpendicular to the plane magnetic field \cite{Tanttu_2019}.

\end{document}